# Stress and Fracture Analysis of a Gravitating Cantilever Beam

Vadapalli Surya Prasanth [0000-0003-2837-4437] and Arun Kumar Singh [0000-0003-1663-9055]

Visvesvaraya National Institute of Technology, Nagpur, India
aksingh@mec.vnit.ac.in

**Abstract.** This research article presents stress distribution and fracture analysis of a cantilever beam considering both continuum and lumped distribution of gravity force. Airy stress function is used to derive two-dimensional stress distribution using power series and boundary conditions [1]. It is concluded that bending stress is the most dominant over other two stresses. However, variation of vertical normal stress does not match between the two systems owing to change in load distribution. Moreover, an expression for energy release rate is also derived by assuming that a crack is present at the top of fixed end of beam and propagates vertically down to result in catastrophic failure. Finally, these results are validated with finite element simulations as well.

## 1    Introduction

Cantilever beams are widely used in real-life structures such as bridges, dams, balconies, mezzanines, staircases etc. These structures are often subjected to self-weight by default, and may result in sudden failure due to crack propagation. Literature survey shows that analytical expression for stresses due to self-weight is very limited. For instance, analytical solution for stresses is available for a gravitating simply supported beam [2]. The expression for stresses is also reported for a gravity dam [3]. Nevertheless, the same for gravitating cantilever beam has not yet been reported in literature. In the present study, we have used a general method to derive the Airy stress function of a cantilever beam under self-weight in view of the method proposed by Neou (1957) [1]. This approach could be used for deriving the solution for stresses in simply supported beam as well [2]. Jayne and Tang (1970) have used Neou's method for determining stresses in anisotropic and orthotropic beams [4].

Fracture analysis of the cantilever beam is also carried out due to its self-weight. It is assumed that a crack is present at the top of the fixed end and that propagates vertically down. In this case too, analytical solution for energy release rate is derived and also matched with numerical simulations.

## 2    Derivation of stresses in a cantilever beam

Airy stress function $\Phi$ of a cantilever beam in Fig.1 owing to gravity force is derived by assuming that gravitational force acts every point of the beam that is, continuum model. Thus, the terms of coefficient $C_{mn}$ of $\Phi$ in view of the double infinite power series [1] as

$$\Phi = \sum_{m=0}^{\infty} \sum_{n=0}^{\infty} C_{mn} \, x^m \, y^n \tag{1}$$

where $m, n$ are non-negative integers. Since gravity $g$ is a conservative force, the potential function V is generally defined [3] as

$$V = -\rho g y \tag{2}$$

where, $\rho$ is the density of the elastic beam. Thus, two-dimensional stresses are often expressed to satisfy the stress-equilibrium equations [3] as

$$\sigma_{xx} = \frac{\partial^2 \Phi}{\partial y^2} + V \, ; \sigma_{yy} = \frac{\partial^2 \Phi}{\partial x^2} + V \, ; \tau_{xy} = -\frac{\partial^2 \Phi}{\partial x \, \partial y} \tag{3}$$

upon substituting Eqs. 1 & 2 in Eq. 3, two-dimensional state of stresses follows:

Normal stress in x-direction,

$$\sigma_{xx} = \sum_{m=0}^{\infty} \sum_{n=2}^{\infty} n(n-1) C_{mn} \, x^m \, y^{n-2} - \rho g y \tag{4.1}$$

Normal stress in y-direction,

$$\sigma_{yy} = \sum_{m=2}^{\infty} \sum_{n=0}^{\infty} m(m-1) C_{mn} \, x^{m-2} \, y^n - \rho g y \tag{4.2}$$

And shear stress in xy-plane,

$$\tau_{xy} = \sum_{m=1}^{\infty} \sum_{n=1}^{\infty} mn \, C_{mn} \, x^{m-1} \, y^{n-1} \tag{4.3}$$

Fig. 1, shows a typical elastic and isotropic cantilever beam, wherein gravitational force is acting along positive-$y$ direction. Following boundary conditions are considered from Fig.1:

at $y = -a$: $\sigma_{yy} = -w$; $\sigma_{xy} = 0$ \hfill (5.1)

at $y = a$: $\sigma_{yy} = 0$; $\sigma_{xy} = 0$ \hfill (5.2)

at $x = 0$: $F_x = \int_{-a}^{a} \sigma_{xx} \, dy = 0$; $M = \int_{-a}^{a} \sigma_{xx} \, y \, dy = 0$;
and $V = \int_{-a}^{a} \tau_{xy} \, dy = 0$ \hfill (5.3)

further at $x = L$: $F_x = \int_{-a}^{a} \sigma_{xx} \, dy = 0$; $M = \int_{-a}^{a} \sigma_{xx} \, y \, dy = -\rho a g L^2$; and $V = \int_{-a}^{a} \tau_{xy} \, dy = -2\rho a g L$ \hfill (5.4)

Noting that $F_x$ is normal force in x-direction, $M$ is bending moment & $V$ is shear force.

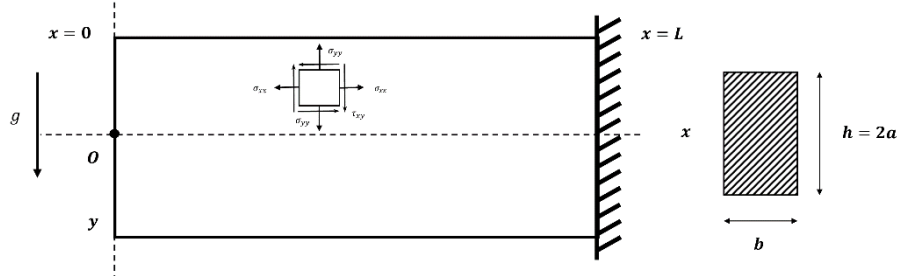

**Fig. 1.** Shows a cantilever beam under self-weight (continuum model) with front view and cross-section view, as well as reference coordinate axes and specifications.

Now two-dimensional stress distribution of the cantilever beam in Fig.1 is obtained by seeking an Airy-stress function $\Phi$ that satisfy both the boundary conditions and the biharmonic equation of compatibility [1] as

$$\nabla^4 \Phi = 0 \qquad (6)$$

solving Eqs. 4, 5 & 6, yields the following Airy stress function

$$\Phi = \frac{3\rho g x^2 y}{4} + \frac{\rho g y^3}{15} - \frac{\rho g x^2 y^3}{4a^2} + \frac{\rho g y^5}{20a^2} \qquad (7.1)$$

while the expressions for two-dimensional stresses in view of Eq.3 are

$$\sigma_{xx} = -\frac{3\rho g y}{5} - \frac{3\rho g x^2 y}{2a^2} + \frac{\rho g y^3}{a^2} \qquad (2.2)$$

$$\sigma_{yy} = \frac{\rho g y}{2} - \frac{\rho g y^3}{2a^2} \qquad (7.3)$$

$$\tau_{xy} = -\frac{3\rho g x}{2} + \frac{3\rho g x y^2}{2a^2} \qquad (7.4)$$

It is significant to note that the present methodology has also enabled us to find the stress distribution for a gravitating simply supported beam as was first reported by Tanimoto (1957) [2].

## 3 Derivation of stresses under GEUDL

Motivated by the derivation of stress equations of the cantilever beam in Fig.1, it becomes interesting to study the distribution of stresses in a cantilever beam in Fig.2 under the gravity equivalent uniformly distributed load (GEUDL) acting at the top (or bottom) of the beam. This study enables one to compare the distribution of stresses between the continuum model (Fig.1) and the GEUDL (lumped) model in Fig.2.

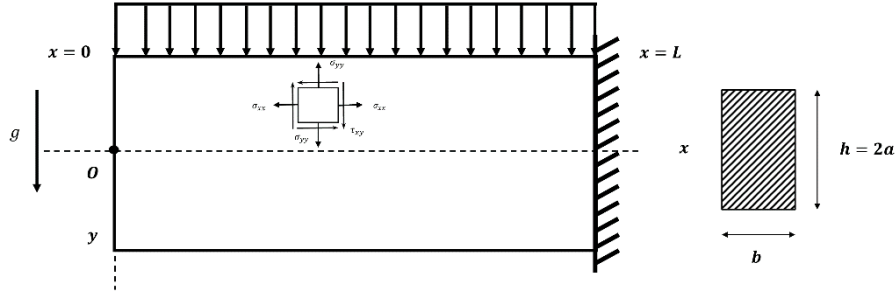

**Fig. 2.** A cantilever beam under GEUDL (lumped model) and also shows front view and cross-section view, with reference coordinate axes.

Let $w\ (= 2a\rho g)$ be load per unit length of the beam in Fig.2, the boundary conditions are used in the solution of Eq.6 as follows

$$at\ y = -a: \sigma_{yy} = -w; \sigma_{xy} = 0 \tag{8.1}$$

$$y = a: \quad \sigma_{yy} = 0; \sigma_{xy} = 0 \tag{8.2}$$

$$x = 0: \quad F_x = \int_{-a}^{a} \sigma_{xx}\,dy = 0;\ M = \int_{-a}^{a} \sigma_{xx}\,y dy = 0;$$
$$V = \int_{-a}^{a} \tau_{xy}\,dy = 0 \tag{8.3}$$

$$x = L: \quad F_x = \int_{-a}^{a} \sigma_{xx}\,dy = 0;\ M = \int_{-a}^{a} \sigma_{xx}\,y dy = -\frac{wL^2}{2};$$
$$and\ at\ V = \int_{-a}^{a} \tau_{xy}\,dy = -wL \tag{8.4}$$

solving Eq. 6 in view of Eq.8, Airy stress function for the beam in Fig.2 is obtained as

$$\Phi = -\frac{wx^2}{4} + \frac{3wx^2y}{8a} - \frac{wx^2y}{20a} - \frac{wx^2y^3}{8a^3} + \frac{wy^5}{40a^3} \tag{9.1}$$

While the equations of stresses are given by

$$\sigma_{xx} = -\frac{3wy}{10a} - \frac{3wx^2y}{4a^3} + \frac{wy^3}{2a^3} \tag{9.2}$$

$$\sigma_{yy} = -\frac{w}{2} + \frac{3wy}{4a} - \frac{wy^3}{4a^3} \tag{9.3}$$

$$\tau_{xy} = \frac{3wx}{4a} + \frac{3wxy^2}{4a^3} \tag{9.4}$$

It is important to note that irrespective of GEUDL acts bottom or top of the surface of the cantilever beam (Fig.2), Airy stress function and corresponding stresses lead to the same expressions as in Eqs.(9.1-4).

## 4    Fracture analysis of the cantilever beam

Crack propagation can be a significant issue in structures subjected to constant loading such as self-weight. High tensile stress at the top corner of the fixed end of cantilever beam (Figs.1&2) could make it particularly susceptible to crack initiation and propagation [5]. Shear force and bending moment due to self-weight of the cantilever beam (Fig.1) are given by the following expressions:

$$V_x = \int_0^x w\, dx = 2ab\rho_s gx \; ; \; M_x = \int_0^x -V_x\, dx = ab\rho_s gx^2 \tag{10}$$

Further total elastic energy stored in the beam is expressed by the following expression [5]:

$$U_E = \int_0^L \left(\frac{M^2}{2EI} + \chi(1+\nu)\frac{T^2}{ES}\right) dx \tag{11}$$

Assuming that cross-section of the cantilever beam to be rectangular (Fig.1 & 2) and shear factor $\chi = 1.20$ [5], Eqs. (10&11) yield to

$$U_E = \int_0^L \left(\frac{(ab\rho_s gx^2)^2}{2EI} + 1.2(1+\nu)\frac{(2ab\rho_s gx)^2}{ES}\right) dx \tag{12}$$

Using the definition of energy release rate [5], in terms of crack length $a_c$ is given by $G = -\frac{\partial U_E}{b\, \partial a_c}$. Hence the analytical solution results in

$$G = \frac{(a\rho_s g)^2 L^3}{E}\left[\frac{3.6L^2}{(h-a_c)^4} + \frac{1.6(1+\nu)}{(h-a_c)^2}\right] \tag{33}$$

Noting that Eq. 13 is also valid for the lumped load in Fig.2. Further, expression of G in Eqs. (12) is also validated numerically using finite element method in next section.

## 5    Results and discussion

### 5.1    Continuum analysis of the beam

Aiming to validate the analytical solutions in Eq. 7, the numerical simulations are carried out for both self-weight (Fig.1) and GEUDL (Fig.2) beams by using Fusion360 software [6]. Assuming that beams are made a typical steel material with elastic modulus E=210 GPa, Poisson's ratio=0.3, density =7800 kg/m$^3$ and g=9.81 m/sec. Results in Fig.3 illustrate a good match of different stresses ($\sigma_{xx}$, $\sigma_{yy}$ & $\tau_{xy}$ vs. depth) between analytical and numerical solutions. Fig.3 also demonstrates that linearly varying bending stress $\sigma_{xx}$ is largest in magnitude than $\sigma_{yy}$ & $\tau_{xy}$. At the same time, variation of $\tau_{xy}$ is parabolic in nature.

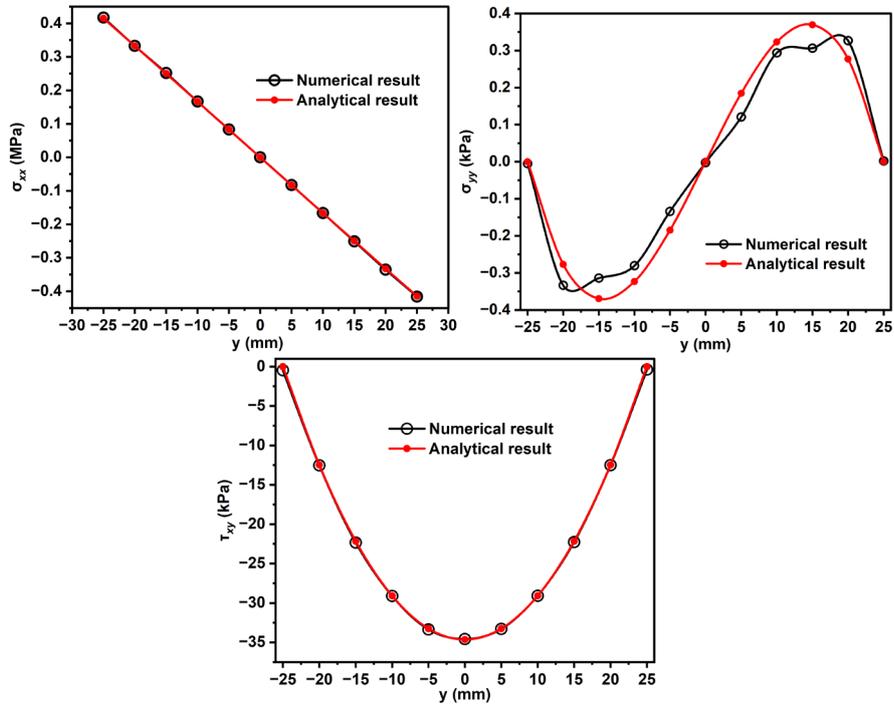

**Fig. 3.** Shows a match between analytical and numerical (FEA) solutions for two-dimensional stresses $\sigma_{xx}, \sigma_{yy}$ & $\tau_{xy}$ of the continuum model.

## 5.2 GEUDL analysis of the beam

We have also investigated stress distribution of the cantilever beam in Fig.2 because of GEUDL on the beam. Results in Fig.4 illustrate a comparison of different stresses($\sigma_{xx}$, $\sigma_{yy}$ & $\tau_{xy}$) between analytical and numerical (FEA) solutions. It is obvious that stresses obtained from two methods are matching quite well.

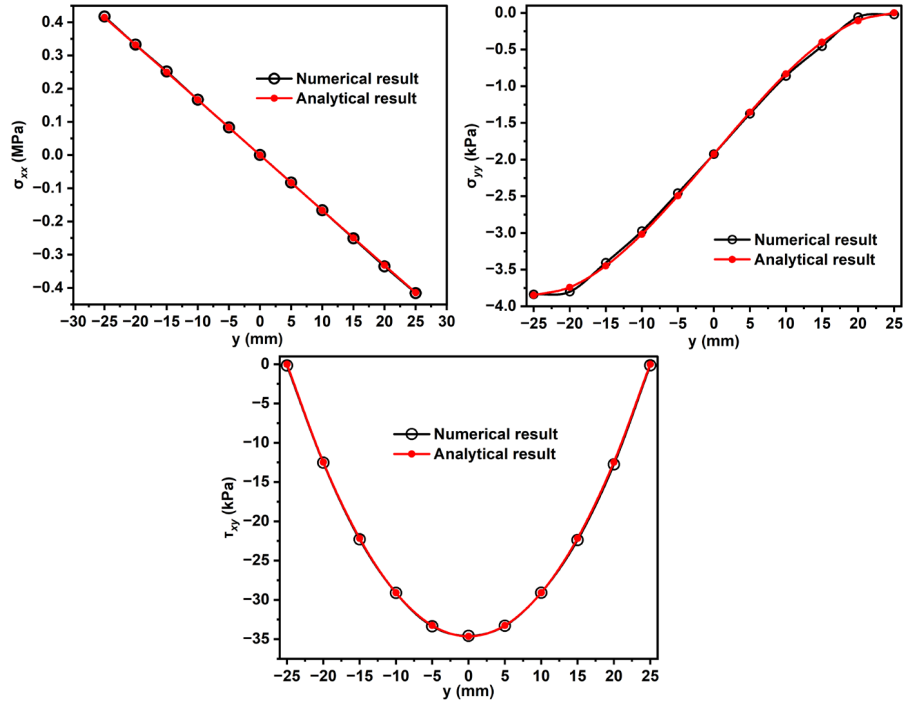

**Fig. 4.** Shows analytical and numerical stresses $\sigma_{xx}, \sigma_{yy}$ & $\tau_{xy}$ for the GEUDL on the beam.

We have also analyzed the stress distribution between continuum and GEUDL on the beam in Figs.1&2. Results in Figs.5&6 demonstrate that magnitude $\sigma_{xx}$ and $\tau_{xy}$ match well, but $\sigma_{yy}$ differs because of distribution between the two beams. Similar observation is also seen for the simply supported beam [3].

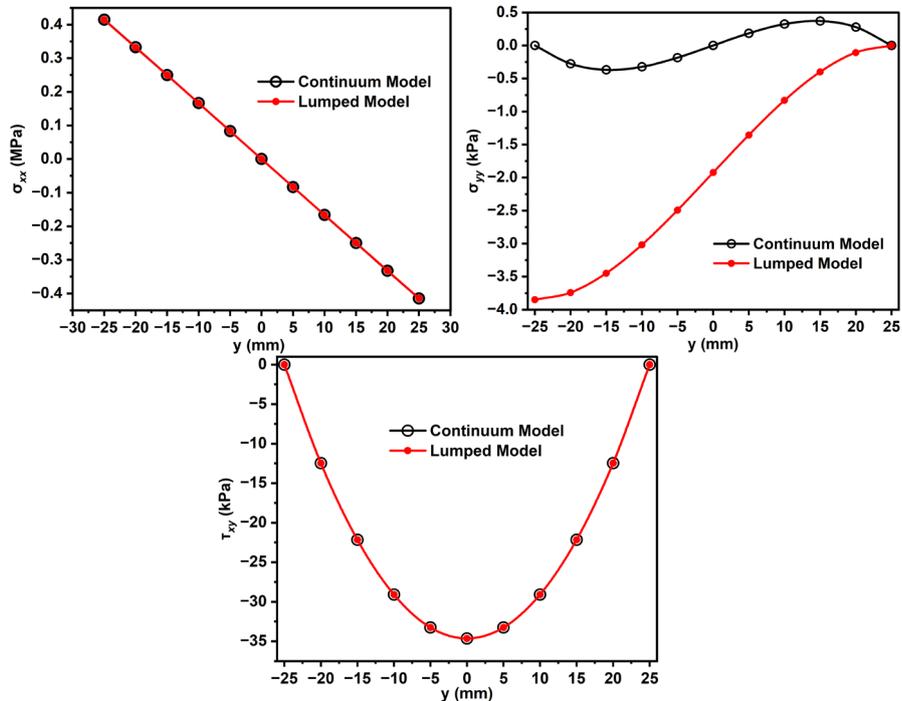

**Fig. 5.** Variation of two-dimensional stresses $\sigma_x, \sigma_y$ & $\tau_{xy}$ vs. thickness y of the cantilever beam under the continuum and GEUDL loading.

### 5.3 Numerical validation of energy release rate

We have also validated the analytical solution for energy release rate G in Eq.13 with finite element analysis (FEA) with ANSYS APDL software [7]. Plane182 element is used for free quad meshing of the beam and that resulted in 4200 mesh elements. Further, it is also assumed that a sharp crack is present at the top of fixed end of the cantilever beam. Node release method was used for deriving the expression for $G$ [8,9]. Result in Fig.6 depicts that although analytical $G$ is quite close to the numerical solution, a correction factor is proposed to match the two solutions more accurately.

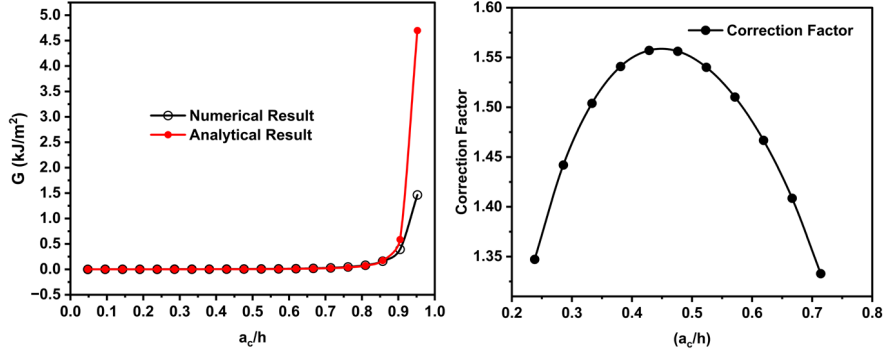

**Fig. 6.** Variation of energy release rate $G$ with relative crack length $(a_c/h)$ and correction factor plots of cantilever beam due to self-weight.

Following correction factor (CF) is included in the expression of G, as following

$$G = \frac{(a\rho_S g)^2 L^3}{E}\left[\frac{3.6L^2}{(h-a_c)^4} + \frac{1.6(1+\nu)}{(h-a_c)^2}\right]\left\{-4.0602\left(\frac{a}{h}\right)^2 + 3.8372\left(\frac{a}{h}\right) + 0.657\right\} \tag{14}$$

Finally, considering fracture toughness of a typical steel material $G_c = 0.7\ kJ.m^{-2}$ [10], while the result in Fig.6, shows the value of $G = 0.9\ kJ.m^{-2}$ at which catastrophic failure of the beam may occur. Hence, concluding that the cantilever beam would fail without any warning once the crack reaches to its critical length.

## 6    Conclusions

Analytical and numerical stress analysis of the cantilever beam due to self-weight reveal that magnitude of bending stress is much higher than remaining two stresses. However, vertical normal stress does not match between continuum and GEUDL beam systems. Stresses obtained from analytical solutions match closely with the numerical solutions. Further, fracture analysis of the cantilever beam is also carried out to derive the expression for energy release rate analytically. At the end, a correction factor is proposed in the energy release rate to match the numerical solution more accurately.

## 7    Declarations

**Conflict of interest**

The authors declare that they have no known competing financial interests or personal relationships that could have appeared to influence the work reported in this paper